# A Leakage-Proof Benchmark and Conformal Selective Triage for Electrohysterogram-Based Preterm Birth Prediction

Sunday A. Adetunji, MD, MPH[1]
[1]*Department of Biostatistics and Epidemiology*
*College of Health, Oregon State University*
Corvallis, USA
**Corresponding author:** Sunday A. Adetunji,
sundayadetunjisa@gmail.com

Rhoda O. Oyewusi, RN, RM[2]
[2]*Department of Nursing & Midwifery, Alifort Hospital Ltd,*
Lagos, Nigeria

*Abstract*—**Background: Preterm birth, defined as delivery before 37 completed gestational weeks, affects approximately 10.6% of live births globally, corresponding to approximately 14.84 million births annually, and remains a major cause of neonatal morbidity and mortality. Electrohysterography (EHG), the noninvasive abdominal recording of uterine myoelectrical activity, has been investigated for preterm-birth risk prediction. However, performance estimates can be biased when multiple segments derived from the same maternal record occur across training and validation partitions, because apparent discrimination may then reflect within-record dependence rather than between-patient prognostic information.**

**Objective: To distinguish segment-level from patient-independent record-grouped validation, establish a patient-independent reference benchmark on the Term-Preterm Electrohysterogram Database, and evaluate class-conditional conformal selective prediction.**

**Methods: All 300 records, including 38 preterm births, were evaluated across three prespecified regimes: segment-level cross-validation, patient-independent record-grouped nested cross-validation, and gestational-age-stratified record-grouped evaluation. A 92-feature engineered EHG representation was modeled using elastic-net logistic regression. Preprocessing, feature standardization, model estimation, Platt calibration, and conformal quantile estimation were confined to training records within each outer fold. Performance was estimated exclusively from held-out out-of-fold record-level predictions using 1,000 record-level bootstrap resamples.**

**Results: Under patient-independent evaluation, the area under the receiver operating characteristic curve (AUROC) was 0.493 (95% CI, 0.467-0.520), the area under the precision-recall curve (AUPRC) was 0.122 (0.095-0.152), and the Brier score was 0.115 (0.094-0.137). AUROC was 0.514 for recordings obtained at or before 26 weeks and 0.469 thereafter. At conformal miscoverage $\alpha$=0.10, marginal coverage was 0.897, abstention was 72.7%, and singleton-prediction accuracy was 0.624.**

**Conclusion: The evaluated engineered EHG representation showed no evidence of discrimination beyond chance when performance was estimated between previously unseen maternal records. More generally, for segmented physiological data, the unit of resampling should correspond to the unit at which predictive performance is intended to generalize. Record-level separation therefore provides a reference condition for distinguishing between-patient prognostic information from dependence among repeated measurements of the same individual. Class-conditional conformal prediction complements this principle by representing uncertainty through set-valued prediction; coverage, abstention, and singleton accuracy should therefore be interpreted jointly when evaluating selective prediction.**

**Keywords— electrohysterography; preterm birth; physiological signal prediction; patient-independent validation; grouped cross-validation; data leakage; conformal prediction; selective prediction; prediction uncertainty.**

## I. INTRODUCTION

Preterm birth (PTB), defined as delivery before 37 completed weeks of gestation, is a major cause of neonatal morbidity and mortality worldwide. The global PTB rate was estimated at 10.6% (uncertainty interval, 9.0%-12.0%) in 2014, corresponding to approximately 14.84 million liveborn preterm neonates annually, with more than 80% of the absolute burden concentrated in South Asia and sub-Saharan Africa [1]. Complications including respiratory distress syndrome, intraventricular hemorrhage, necrotizing enterocolitis, sepsis, and cerebral palsy account for a substantial proportion of neonatal and under-five mortality; survivors have elevated long-term neurocognitive and cardiometabolic risk [1], [2]. Because preventive and preparatory interventions can reduce adverse outcomes when applied prospectively, accurate early identification of pregnancies at increased risk has clinical and public-health value.

Electrohysterography (EHG) is the noninvasive recording of uterine myoelectrical activity with abdominal surface electrodes. It has been proposed as a physiologically grounded, low-cost PTB risk-stratification modality since Fele-Zorz et al. showed differences in spectral and nonlinear EHG descriptors between term and preterm groups [3], [4]. The public PhysioNet Term-Preterm EHG Database (TPEHGDB) [5] has accelerated this research direction. Nevertheless, published EHG-based PTB studies have reported areas under the receiver operating characteristic curve (AUROCs) above 0.80-0.97, whereas performance commonly declines under rigorous subject-independent replication [6], [7].

A principal threat is evaluation leakage. Standard preprocessing pipelines may segment each approximately 30-minute recording into multiple windows and then split at the window rather than the record level, allowing models to exploit record-specific structure instead of transportable physiology [8], [9]. Similar biomedical prediction studies have documented marked performance inflation when correlated observations from the same subject cross validation partitions [8], [9]. Under prediction-model risk-of-bias principles, failure to preserve subject independence is a high-risk analysis feature [17].

Even after leakage is removed, discrimination alone incompletely characterizes clinical utility under class imbalance and asymmetric misclassification costs. Selective prediction permits a model to abstain from uncertain cases while routing those cases to clinical assessment [10], [11]. Conformal prediction provides a distribution-free framework for set-valued prediction with finite-sample coverage under exchangeability [12]-[14]. This study therefore had three objectives: (1) formalize the distinction between naive segment-level and patient-independent record-grouped evaluation; (2) establish a reproducible leakage-proof benchmark aligned with contemporary prediction-model reporting and risk-of-bias guidance [15]-[17], [22]; and (3) evaluate label-conditional conformal selective triage as an explicit abstention mechanism.

## II. METHODS

### *A. Data Source, Cohort Assembly, and Ethics*

All analyses used TPEHGDB version 1.0.1 (PhysioNet; doi:10.13026/C2FW2V), a publicly available, fully de-identified resource comprising 300 unique EHG recordings from singleton pregnancies at the University Medical Centre Ljubljana, Slovenia [5]. The analytic cohort comprised 38 preterm records (12.7%) and 262 term records (87.3%). Each unique record was the atomic unit of inference in all leakage-proof analyses. This secondary analysis used a fully de-identified public dataset; no additional institutional review board review or informed consent was required.

### *B. Outcome Definition and Temporal Stratification*

PTB was defined as delivery gestational age (GA) below 37 weeks and term birth as delivery GA of at least 37 weeks, consistent with the TPEHGDB label convention [5]. A prespecified temporal boundary classified recording GA at or before 26 weeks as early (n = 162; 19 preterm) and recording GA after 26 weeks as late (n = 138; 19 preterm). Delivery GA and recording GA were used only for outcome labeling and stratum assignment and were excluded from predictor inputs.

### *C. Feature Representation and Fold-Pure Preprocessing*

Precomputed TPEHGDB features [5] under three band-pass filters (0.08-4.0 Hz, 0.3-3.0 Hz, and 0.3-4.0 Hz) across three bipolar EHG channels included root mean square (RMS) amplitude, median frequency, peak frequency, sample entropy (SampEn), and across-channel summary statistics. Cross-filter differences captured inter-band changes, yielding a fixed 92-dimensional fused feature vector per record. Median imputation and interquartile-range-based robust scaling were estimated exclusively in the training partition of each outer fold and then applied unchanged to held-out data through Scikit-learn pipeline objects [18].

### *D. Experimental Design: Three-Regime Leakage Benchmark*

*Regime 1 - naive segment-level cross-validation:* Segments were treated as independent samples without record grouping. This leakage-prone reference reproduces the common protocol in which windows from the same record may cross partitions [8], [9].

*Regime 2 - patient-independent record-grouped nested cross-validation:* The record identifier was the exclusive grouping variable. All data from a record remained in one fold. Within each outer training fold, records were divided into mutually disjoint model-fit, Platt-calibration, and conformal-quantile subsets. Strictly out-of-fold (OOF) predictions were generated for held-out records with no exposure to any fitting, preprocessing, calibration, or conformal-estimation stage.

*Regime 3 - temporally stratified record-grouped evaluation:* Regime 2 was repeated with performance reported separately for early and late recording strata to characterize gestational covariate shift without introducing stratum-level leakage. Primary inferential claims were based on Regimes 2 and 3 [8], [9], [15], [17].

### *E. Model, Calibration, and Conformal Selective Triage*

Elastic-net regularized logistic regression with inverse-frequency class weighting was selected for interpretability and stability under small-sample class imbalance. Hyperparameters were tuned within training folds. Platt sigmoid calibration [19] was fitted on a disjoint within-fold calibration subset. Class-conditional conformalization [12]-[14] used nonconformity score $s(x, y) = 1 - \hat{p}(y \mid x)$ and separate class-conditional quantiles estimated on the disjoint conformal subset at miscoverage level 0.10. The prediction set was defined as the set of labels whose nonconformity scores did not exceed the corresponding class-conditional quantile.

$$P(Y \in \hat{C}(X)) \geq 1 - \alpha, \quad \alpha = 0.10 \tag{1}$$

where Y denotes the observed delivery class, $\hat{C}(x)$ is the conformal prediction set, and the nominal target coverage is 0.90. The triage actions were {0}, predict term; {1}, predict preterm; and {0,1}, defer or uncertain and escalate to clinical assessment [10], [11].

### *F. Performance Metrics and Uncertainty Quantification*

Discrimination was summarized by area under the receiver operating characteristic curve (AUROC) and area under the precision-recall curve (AUPRC), with AUPRC interpreted relative to the observed prevalence of 0.127. Probabilistic accuracy was summarized by the Brier score. Selective-prediction endpoints were marginal coverage, abstention rate, singleton rate, and singleton-conditional accuracy. Uncertainty was quantified with 1,000 nonparametric record-level bootstrap resamples and percentile 95% confidence intervals [20]. Reporting followed contemporary prediction-model reporting and risk-of-bias guidance [15]-[17], [22].

## III. RESULTS

### *A. Cohort Composition and Gestational Timing*

The cohort comprised 300 unique records: 38 preterm (12.7%) and 262 term (87.3%). The prespecified strata included 162 recordings at or before 26 weeks (19 preterm, 11.7%) and 138 after 26 weeks (19 preterm, 13.8%). Delivery GA differed by outcome, whereas recording GA was determined by the temporal stratum. Both variables were excluded from predictor inputs (Table I; Fig. 1).

TABLE I. COHORT COMPOSITION BY RECORDING GESTATIONAL STRATUM AND DELIVERY OUTCOME

| Stratum | Outcome | N | Delivery GA, weeks (mean ± SD) | Recording GA, weeks (mean ± SD) |
|---|---|---|---|---|
| ≤26 weeks | Preterm (<37 weeks) | 19 | 34.2 ± 2.7 | 23.4 ± 0.7 |
| ≤26 weeks | Term (≥37 weeks) | 143 | 39.7 ± 1.1 | 23.1 ± 0.8 |
| >26 weeks | Preterm (<37 weeks) | 19 | 34.7 ± 2.0 | 30.6 ± 1.1 |
| >26 weeks | Term (≥37 weeks) | 119 | 39.6 ± 1.1 | 31.2 ± 1.0 |

*Values are mean ± standard deviation. GA indicates gestational age; SD, standard deviation. Delivery and recording GA were used only for labeling or stratification and were excluded from predictors.*

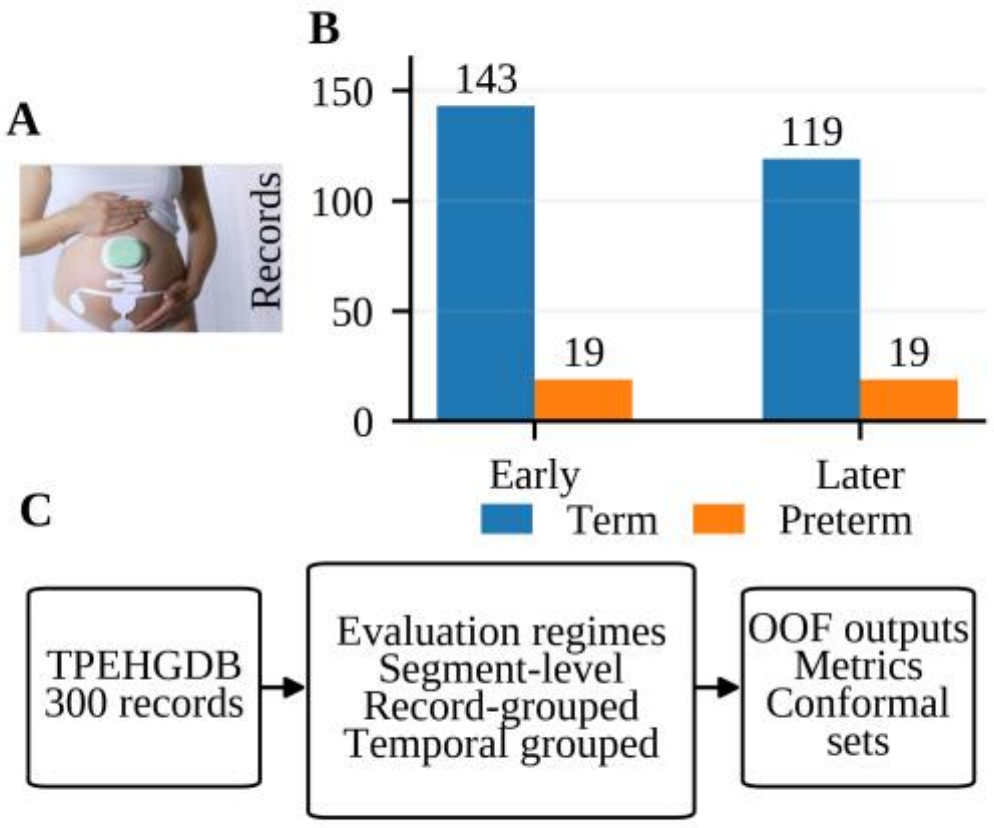

Fig. 1. Clinical context, cohort composition, and leakage-proof evaluation architecture. (A) Representative wearable abdominal EHG monitoring context; the image was illustrative only and was not used for model training, validation, or inference [21]. (B) Cohort composition by recording-time stratum and outcome. (C) Three-regime benchmark showing naive segment-level cross-validation, patient-independent record-grouped nested cross-validation, and temporally stratified record-grouped evaluation. CV indicates cross-validation; OOF, out-of-fold.

### *B. Feature-Space Structure*

A two-dimensional principal component analysis (PCA) projection of the standardized 92-dimensional fused feature representation showed substantial overlap between term and preterm records. The PCA display was used for visualization only and was not part of the predictive pipeline.

### *C. Primary Performance Under Leakage-Proof Evaluation*

Under strict patient-independent, record-grouped OOF evaluation, overall discrimination was near chance. The elastic-net logistic model with within-fold Platt calibration and label-conditional conformalization achieved AUROC 0.493 (95% confidence interval, 0.467-0.520), AUPRC 0.122 (0.095-0.152), and Brier score 0.115 (0.094-0.137). The AUPRC was close to the observed prevalence baseline of 0.127, and the predicted-risk distributions showed substantial overlap between term and preterm records (Table II; Fig. 2).

TABLE II. PRIMARY LEAKAGE-PROOF OUT-OF-FOLD PERFORMANCE

| Quantity | Estimate |
|---|---|
| Model | Elastic-net logistic regression + Platt calibration + conformal prediction |
| Pooled OOF prediction instances | 3,000 |
| Prevalence | 12.7% |
| AUROC (95% CI) | 0.493 (0.467-0.520) |
| AUPRC (95% CI) | 0.122 (0.095-0.152) |
| Brier score (95% CI) | 0.115 (0.094-0.137) |

*OOF indicates out-of-fold; CI, confidence interval. Pooled OOF instances arise from repeated record-grouped cross-validation and are not unique records. Confidence intervals used 1,000 record-level bootstrap resamples.*

### *D. Stratified Performance by Recording Gestational Age*

Performance differed directionally across gestational recording strata: AUROC was 0.514 for recordings at or before 26 weeks and 0.469 for later recordings; Brier scores were 0.108 and 0.124, respectively. This pattern is consistent with temporal covariate shift but does not constitute a formal domain-shift test.

TABLE III. OUT-OF-FOLD PERFORMANCE BY RECORDING GESTATIONAL STRATUM

| Metric | ≤26 weeks | >26 weeks |
|---|---|---|
| OOF instances | 1,620 | 1,380 |
| Prevalence | 11.7% | 13.8% |
| AUROC | 0.514 | 0.469 |
| AUPRC | 0.123 | 0.123 |
| Brier score | 0.108 | 0.124 |

*Metrics were calculated from record-grouped OOF predictions within each prespecified stratum. No formal distribution-shift test was applied.*

### *E. Conformal Selective Triage*

At miscoverage level 0.10, marginal coverage was 0.897, the abstention rate was 72.7%, the singleton rate was 27.3%, and singleton-conditional accuracy was 0.624 (Table IV; Fig. 3). Uncertain cases were explicitly deferred to clinical assessment.

TABLE IV. CONFORMAL SELECTIVE-PREDICTION METRICS AT MISCOVERAGE LEVEL 0.10

| Metric | Value |
|---|---|
| Marginal coverage | 0.897 (target 0.90) |
| Abstention rate, set {0,1} | 72.7% |
| Singleton rate | 27.3% |
| Singleton-conditional accuracy | 0.624 (4.9 x prevalence) |

*Coverage is the proportion of cases for which the observed label is contained in the conformal prediction set.*

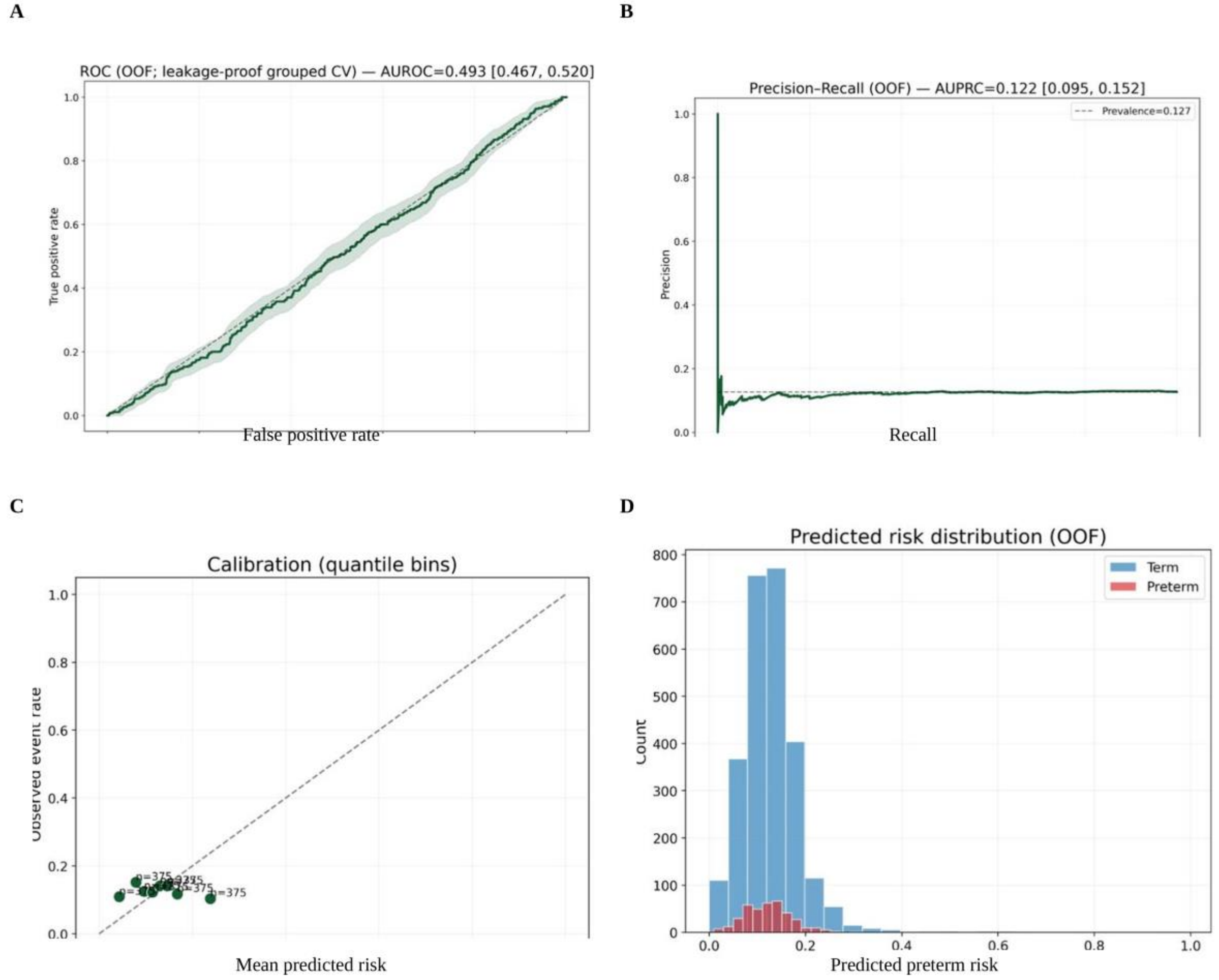

Fig. 2. Primary leakage-proof model performance from record-grouped out-of-fold predictions. (A) Receiver operating characteristic curve. (B) Precision-recall curve with the observed preterm-birth prevalence as the horizontal reference. (C) Quantile-binned calibration. (D) Predicted-risk distributions for term and preterm records.

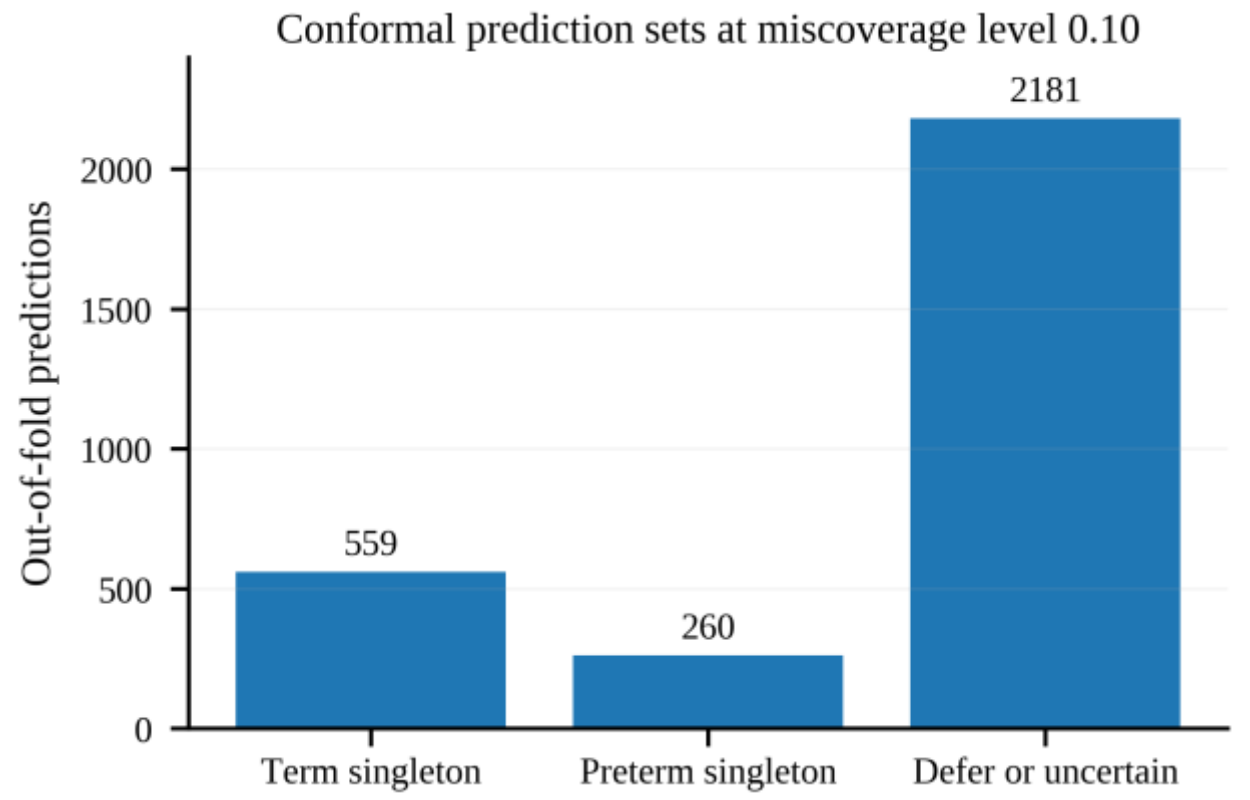

Fig. 3. Conformal selective triage: frequency of singleton term, singleton preterm, and defer or uncertain prediction sets at miscoverage level 0.10. Counts reflect pooled OOF predictions across repeated cross-validation.

*F. Feature Interpretability and Robustness*

Feature-stability rankings were led by peak-frequency minimum in the 0.3-4.0 Hz and 0.3-3.0 Hz bands and by the cross-filter RMS difference. For the cross-filter RMS feature, the mean signed coefficient was +0.576 (SD 0.460) and the mean absolute coefficient was 0.636. Across feature-view sensitivity analyses, AUROC ranged from 0.483 to 0.500; across random feature-dropout analyses, AUROC ranged from 0.478 to 0.504. No evaluated sensitivity condition produced a material increase in discrimination (Tables V-VII).

TABLE V. TOP FEATURES BY STANDARDIZED ELASTIC-NET COEFFICIENT STABILITY

| **Feature** | **Mean coef.** | **SD** | **Mean abs. coef.** | **Nonzero folds** |
|---|---|---|---|---|
| 0.3-4.0 Hz Fpeak_min | -0.757 | 0.291 | 0.757 | 98% |
| 0.3-3.0 Hz Fpeak_min | -0.730 | 0.285 | 0.730 | 98% |
| XFILT ΔRMS (0340-0330), mean | +0.576 | 0.460 | 0.636 | 100% |
| 0.3-3.0 Hz SampEn channel 3 | -0.527 | 0.368 | 0.557 | 100% |
| 0.3-3.0 Hz Fmed channel 3 | -0.476 | 0.466 | 0.548 | 94% |

*Top 5 of 92 features are shown. XFILT indicates cross-filter difference features; SampEn, sample entropy.*

TABLE VI. FEATURE-VIEW SENSITIVITY ANALYSIS

| **Feature view** | **Dim.** | **AUROC** | **AUPRC** | **Brier** |
|---|---|---|---|---|
| 0.08-4.0 Hz only | 28 | 0.500 | 0.127 | 0.116 |
| 0.3-3.0 Hz only | 28 | 0.499 | 0.128 | 0.117 |
| 0.3-4.0 Hz only | 28 | 0.491 | 0.123 | 0.119 |
| XFILT only | 8 | 0.483 | 0.119 | 0.116 |
| Fused all + derived | 92 | 0.493 | 0.125 | 0.117 |

TABLE VII. FEATURE-DROPOUT ROBUSTNESS

| **Random dropout fraction** | **AUROC** | **AUPRC** | **Brier** |
|---|---|---|---|
| 0% (robustness comparator) | 0.478 | 0.120 | 0.117 |
| 10% | 0.504 | 0.129 | 0.117 |
| 30% | 0.494 | 0.123 | 0.116 |

*Random dropout was applied within folds. The 0% row is the comparator within this robustness analysis; Table II reports the primary model estimate.*

# IV. DISCUSSION

## A. Evidentiary Recalibration Under Subject-Independent Evaluation

The central finding was that the evaluated engineered EHG representation provided no evidence of discrimination beyond chance under patient-independent, record-grouped evaluation (AUROC 0.493). This finding shifts the evidentiary question from whether an EHG model can achieve high discrimination under a particular resampling scheme to whether that discrimination persists when all observations from the same maternal record are confined to a single validation partition. Prior TPEHGDB studies have reported substantially higher performance, including AUROCs of approximately 0.74–0.78 for a deep bidirectional long short-term memory model evaluated with record grouping [7]. Differences between those estimates and the present benchmark may reflect differences in signal representation, model class, preprocessing, partitioning, or other sources of analysis dependence and should therefore be resolved through directly comparable subject-independent evaluations.

The principal contribution of this study is the establishment of a patient-level evidentiary benchmark in which preprocessing, model estimation, probability calibration, and conformal quantile estimation are all confined to training data and performance is evaluated exclusively from held-out out-of-fold predictions. In segmented physiological data, this distinction is fundamental: independence of signal windows is not equivalent to independence of patients. Consequently, performance estimates obtained after preserving patient-level independence more directly characterize between-patient prognostic information rather than within-record similarity. The present benchmark therefore provides a reproducible reference against which future EHG representations, raw-waveform models, multimodal predictors, and prospective cohorts can be evaluated under the same inferential conditions [15]–[17], [22]. Progress in EHG-based preterm-birth prediction should be demonstrated by reproducible improvement under patient-independent validation, not by increasing the number or granularity of correlated observations derived from the same physiological record.

## B. Leakage Mechanisms and Reproducibility Implications

The three-regime evaluation design distinguishes segment-level partitioning from patient-independent record-grouped validation within the same EHG dataset. This distinction is inferential rather than procedural: when multiple observations originate from the same individual, the unit of resampling must correspond to the unit at which predictive performance is intended to generalize. Tougui et al. [8] demonstrated that allowing correlated observations to occur across validation partitions can materially inflate diagnostic performance estimates, while Varma and Simon [9] showed that data-dependent model selection performed outside the validation hierarchy can bias estimates of prediction error. The present study extends these principles to obstetric electrophysiology. All segments derived from a maternal record were assigned to the same outer partition, and every data-dependent operation—including preprocessing, feature standardization, model estimation, probability calibration, and conformal estimation—was performed without access to the corresponding held-out records. Under this design, out-of-fold performance estimates target discrimination between previously unseen maternal records rather than discrimination among correlated segments derived from records already represented during model development. More generally, for clustered biomedical data, valid external prediction error requires separation at the highest observational level shared by correlated measurements; otherwise, the estimated quantity can reflect within-subject dependence rather than between-subject prognostic information.

## C. Conformal Triage and Explicit Deferral

Class-conditional conformal prediction extends evaluation beyond forced binary classification by allowing the prediction set to reflect the degree of uncertainty supported by the observed data. Under exchangeability, the procedure provides finite-sample marginal coverage guarantees [12]–[14]. In the present analysis, marginal coverage approached the nominal 0.90 level, while 72.7% of predictions contained both outcome classes and 27.3% were singletons. These quantities should be interpreted jointly: coverage describes whether the observed outcome is contained within the prediction set, whereas abstention quantifies how often the available information is insufficient to support a unique class assignment. A procedure can therefore attain nominal coverage while remaining noncommittal for a large proportion of observations. This distinction is particularly important in clinical prediction, where uncertainty should be represented rather than concealed by compulsory classification. Accordingly, conformal performance should be reported not by coverage alone, but together with prediction-set size, abstention frequency, and accuracy conditional on singleton prediction. This provides a statistically explicit basis for distinguishing predictions that support a unique classification from those that require additional clinical information or assessment

## D. Gestational Domain Shift and Future Evaluation

Discrimination was lower for recordings obtained after 26 weeks than for those obtained at or before 26 weeks (AUROC 0.469 versus 0.514). Although these stratum-specific estimates do not establish a distributional shift, their direction is biologically compatible with established gestational changes in uterine electrophysiology [3], [4]. More generally, gestational age should not be treated solely as a baseline covariate when the measured physiological process itself evolves during pregnancy. For longitudinally changing biosignals, predictive validity may depend on the developmental interval in which the signal is acquired, such that performance averaged across gestation can obscure clinically relevant variation in between-patient prognostic information. Gestational time therefore constitutes an explicit dimension of model transportability that should be examined through prespecified temporal strata, standardized acquisition, and prospective external validation. Future EHG studies should combine multicenter physiological recordings with raw-waveform and multimodal models while preserving patient-level separation throughout model development and evaluating whether predictive information remains stable, emerges, or attenuates across gestation.

### *E. Limitations*

The analysis was limited to precomputed engineered EHG features from a single-center public dataset comprising 300 records, including 38 preterm births, which constrained statistical power, precision, and external transportability. This limitation was addressed by defining the maternal record as the unit of analysis, using record-grouped data partitioning, estimating all preprocessing parameters within training folds, and evaluating performance exclusively from held-out out-of-fold predictions, thereby preventing correlated signal segments from inflating the effective sample size or producing information leakage. Accordingly, the findings characterize the evaluated engineered-feature representation and should not be interpreted as an upper bound on the prognostic information contained in raw EHG waveforms. Direct waveform analysis and multimodal prediction incorporating complementary obstetric measurements may produce different estimates and should be evaluated using the same subject-independent validation procedures. The small number of preterm births also reduced precision; this was addressed using 1,000 record-level bootstrap resamples with confidence intervals, although sampling variability remains substantial with 38 events. Minority-class sparsity may affect estimation of class-conditional conformal quantiles; separate outcome-specific calibration distributions were therefore used, but larger independent cohorts are required to assess the reproducibility of preterm-class coverage. The gestational-stratum analysis was prespecified and conducted using record-grouped evaluation; however, because no formal statistical test of distributional heterogeneity was performed, the observed between-stratum differences should be interpreted as hypothesis-generating evidence of gestational variation rather than proof of temporal distribution shift. Finally, uncertain prediction sets were designated for clinician assessment rather than forced classification; prospective multicenter studies are still required to evaluate calibration over time, referral frequency, clinical workflow integration, diagnostic consequences, and maternal and neonatal outcomes.

## V. CONCLUSION

Under strictly patient-independent evaluation, the engineered EHG representation showed no evidence of discrimination beyond chance, with an AUROC of 0.493 and an AUPRC of 0.122, closely approximating the preterm-birth prevalence of 0.127. The central methodological finding is that, in segmented physiological recordings, the validity of predictive performance depends on preserving the independence of the patient—not merely the independence of the signal segment. Record-level partitioning, fold-specific preprocessing, and complete separation of model fitting, calibration, and uncertainty estimation are therefore essential for estimating between-patient prognostic performance without contamination from within-record dependence. The resulting estimates provide a leakage-controlled reference benchmark for future EHG studies. Class-conditional conformal prediction further provides a formal means of distinguishing cases for which a single prediction can be issued from those for which uncertainty warrants abstention. For EHG-based preterm-birth prediction, methodological progress should therefore be measured by reproducible gains in patient-level prognostic information under fully separated validation, rather than by higher discrimination obtained from increasingly granular resampling of the same physiological record. This distinction provides a reproducible basis for determining whether future EHG methods identify clinically transportable physiological information or reproduce dependence already present within the observed recordings.

## ACKNOWLEDGMENT

The authors acknowledge PhysioNet and the original TPEHGDB contributors for maintaining the public resource used in this study [5]. No external funding was received. The authors declare no competing interests. The analysis used only public, de-identified data and involved no new participant recruitment or intervention. Data are available as TPEHGDB [5], and reproducibility materials are available at http://github.com/drsunday-ade/leakage-proof-benchmark-for-preterm.